\documentclass[letterpaper,aps,prl,showpacs,floats,nofootinbib,reprint,lengthcheck,superscriptaddress]{revtex4-1}
\usepackage[utf8]{inputenc}
\usepackage[T1]{fontenc}
\usepackage{amsmath,amssymb}

\usepackage{graphicx,color}
\usepackage{hyperref}
\usepackage{microtype}

\newcommand{\titlestr}{Extraordinary exciton conductance induced by strong coupling}

\hypersetup{colorlinks,hypertexnames=true,pdftitle={\titlestr},linkcolor=blue,citecolor=blue,urlcolor=blue}

\newcommand{\LL}{\mathcal{L}}
\DeclareMathOperator{\Tr}{Tr}

\usepackage{multirow}
\def\svdots{\vbox{\baselineskip=3pt \lineskiplimit=0pt \kern3pt \hbox{.}\hbox{.}\hbox{.}}}
\def\sddots{\lower1pt\hbox{$\smash\ddots$}}
\newcommand{\trvdots}{\multirow{2}{*}{\vbox{\baselineskip=3.5pt \lineskiplimit=0pt \kern3pt \hbox{.}\hbox{.}\hbox{.}\hbox{.}\hbox{.}\hbox{.}}}}

\begin{document}
\title{\titlestr}
\author{Johannes Feist}
\email{johannes.feist@uam.es}
\affiliation{Departamento de F\'isica Te\'orica de la Materia Condensada
  and Condensed Matter Physics Center (IFIMAC),
  Universidad Aut\'onoma de Madrid, E-28049 Madrid, Spain}
\author{Francisco J. Garcia-Vidal}
\email{fj.garcia@uam.es}
\affiliation{Departamento de F\'isica Te\'orica de la Materia Condensada
  and Condensed Matter Physics Center (IFIMAC),
  Universidad Aut\'onoma de Madrid, E-28049 Madrid, Spain}
\affiliation{Donostia International Physics Center (DIPC),
  E-20018 Donostia/San Sebastian, Spain}

\date{\today}
\pacs{71.35.-y, 05.60.Gg, 71.36.+c, 81.05.Fb}

\begin{abstract}
  We demonstrate that exciton conductance in organic materials
  can be enhanced by several orders of magnitude when the molecules
  are strongly coupled to an electromagnetic mode. Using a 1D
  model system, we show how the formation of a collective polaritonic
  mode allows excitons to bypass the disordered array of molecules and
  jump directly from one end of the structure to the other. This
  finding could have important implications in the fields of exciton
  transistors, heat transport, photosynthesis, and biological systems
  in which exciton transport plays a key role.
\end{abstract}

\maketitle

The transport of excitons (bound electron-hole pairs) is a fundamental
process that plays a crucial rule both in natural phenomena such as
photosynthesis, where energy has to be transported to a reaction
center \cite{Scholes2011,Engel2007,Lee2007}, and in artificial devices
such as excitonic transistors \cite{High2008,Saikin2013} or organic
solar cells, whose power conversion efficiency can be improved
significantly when the exciton diffusion length is increased
\cite{Menke2013}. Similarly, understanding and manipulating the role
of excitons in heat transport has become an active field of research,
with possible applications ranging from thermoelectric effects to
heat-voltage converters, to nanoscale refrigerators, and even thermal
logic gates (cf.~\cite{Dubi2011} and references therein). The exciton
transport efficiency depends on a wide range of factors with such
surprising features as the occurrence of noise-assisted transport
\cite{Plenio2008,Caruso2009,Moix2013}. Pioneering works have even
suggested that coherent transport can play an important role in
biological systems \cite{Engel2007,Lee2007,Huelga2013}. However, most
systems composed of organic molecules are disordered and possess
relatively large dissipation and dephasing rates, such that exciton
transport typically becomes diffusive over long distances
\cite{Akselrod2014}.

An intriguing possibility to modify exciton properties is by strong
coupling to an electromagnetic (EM) mode, forming so-called polaritons
(hybrid light-matter states). This is achieved when the Rabi
frequency, i.e., the energy exchange rate between exciton and EM
modes, becomes faster than the decay and/or decoherence rates of
either constituent. Polaritons combine the properties of their
constituents, in particular, mutual interactions and low effective
masses, enabling new applications such as polariton condensation in
semiconductors \cite{Kasprzak2006,*Balili2007} and organic materials
\cite{Kena-Cohen2010,*Plumhof2013}, the modification of molecular
chemistry \cite{Hutchison2012} and work functions
\cite{Hutchison2013}, or the transfer of excitation between different
molecular species \cite{Coles2014}. Because of the large dipole moments
and high densities, organic materials support large Rabi splittings
\cite{Lidzey1998,Schwartz2011,Kena-Cohen2013}, and can also be
strongly coupled to surface plasmon polaritons
\cite{Bellessa2004,Dintinger2005,Hakala2009,Vasa2010,Schwartz2011}. The
dispersion relation can then be tuned to achieve a further reduction
of the effective mass \cite{Rodriguez2013}.

Very recently, an increase of the \emph{electrical} conductance of an
organic material was shown under strong coupling of the excitons to a
cavity mode \cite{Orgiu2014}. Inspired by this result, we demonstrate
in this Letter that through strong coupling to an electromagnetic
mode, i.e., the creation of polaritonic states, the exciton transport
efficiency can be improved by many orders of magnitude. The strong
coupling allows the excitons to bypass the disordered organic system,
preventing localization and leading to dramatically improved energy
transport properties. We note that while we focus on organic molecules
in the following, the results can readily be generalized to other
systems such as quantum dots and Rydberg atoms, or even chains of
trapped ions, which offer a high degree of controllability
\cite{Haze2012,Ramm2014}.

\begin{figure}[tbp]
\includegraphics[width=\linewidth]{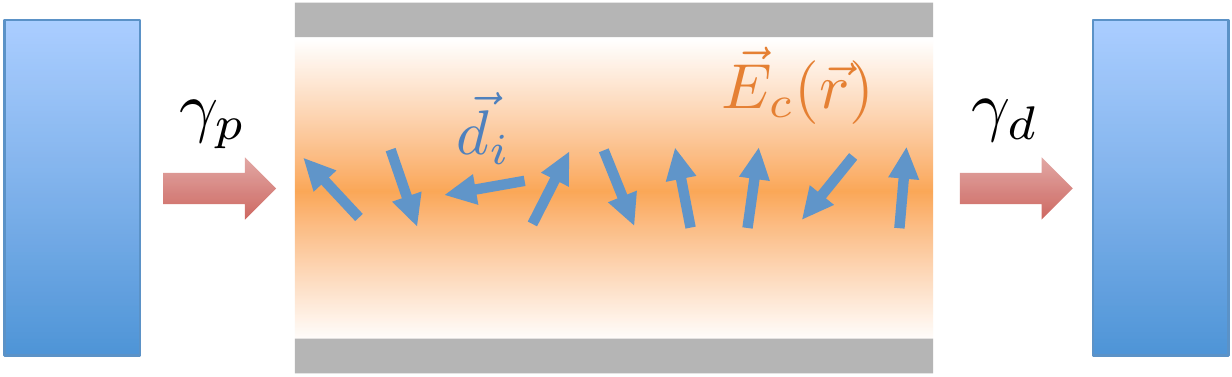}
\caption{Sketch of the model system. A 1D chain of (possibly
  disordered) quantum emitters with dipole moments $\vec d_i$ inside a
  cavity with cavity mode $\vec E_c(\vec r)$. Excitons are pumped into
  the system from the left reservoir with rate $\gamma_p$. The exciton
  current is measured by the excitons reaching the sink reservoir on
  the right, coupled through incoherent decay of the last emitter with
  rate $\gamma_d$.}
\label{fig:system}
\end{figure}

We focus on a model system that captures the essential physics: A 1D
chain of two-level emitters inside a cavity (see
\autoref{fig:system}). The emitter dipole transition is coupled to the
single cavity mode, and, additionally, induces Coulombic dipole-dipole
interaction between the emitters. The effect of internal (e.g.,
rovibrational or phononic) and external environment modes is taken
into account through effective dephasing and nonradiative decay rates
modeled using a master equation of Lindblad form. The system
Hamiltonian $H$ in the rotating wave approximation (setting
$\hbar=1$ here and in the following) is then
\begin{multline}\label{eq:hamiltonian}
H = \omega_c a^\dag a + \sum_i \omega_m \sigma_i^+ \sigma_i^- 
+ \sum_i g_i (a^\dag\sigma_i^- + a\sigma_i^+) \\
+ \sum_{i,j} V^{dd}_{ij} (\sigma_i^+ \sigma_j^- + \sigma_j^+\sigma_i^-) \,,
\end{multline}
where $a$ is the bosonic annihilation operator of the cavity mode with
energy $\omega_c$ and electric field $\vec E_c(\vec r)$. The molecular
excitons of energy $\omega_m$ are created and destroyed by the
fermionic operators $\sigma_i^+$ and $\sigma_i^-$. Molecule $i$ is
characterized by its position $\vec r_i$ and dipole moment $\vec d_i$,
giving the cavity-molecule interaction $g_i = -\vec d_i \cdot
\vec E_c(\vec r_i)$. The dipole-dipole interaction (in the quasistatic
limit) is
\begin{equation}
V^{dd}_{ij} = \frac{\vec d_i \cdot \vec d_j -
3 (\vec d_i \cdot \hat R_{ij}) (\vec d_j \cdot \hat R_{ij}) }{4 \pi \epsilon_0 R_{ij}^3},
\end{equation}
with $R_{ij} = |\vec r_i - \vec r_j|$ and $\hat R_{ij} = (\vec r_i - \vec r_j)/R_{ij}$.

The system dynamics is described by a Lindblad master equation for
the density matrix
\begin{equation}\label{eq:master}
\dot\rho = -i [H,\rho] + \sum_{\alpha=i,c} \LL_\alpha[\rho],
\end{equation}
where $\alpha$ runs over all molecules as well as the cavity mode.
The superoperators $\LL_\alpha$ describe decay and dephasing:
\begin{align}\label{eq:superops}
\LL_i &= \gamma_d L_{\sigma_i^-} + \gamma_\phi L_{\sigma_i^+\sigma_i^-},\\
\LL_c &= \kappa L_{a},
\end{align}
where $L_{c}[\rho] = \frac12 (2 c\rho c^\dag - \{c^\dag c, \rho\})$ is
the standard form for Lindblad superoperators. The total molecule
decay rate $\gamma_d$ is given by $\gamma_d=\gamma_r+\gamma_{nr}$,
with $\gamma_r$ and $\gamma_{nr}$ the radiative and nonradiative decay
rates, while $\gamma_\phi$ is the dephasing rate. The decay rate
$\kappa$ of the cavity photons is dominated by leakage through the
mirrors. For later reference, we also define the total molecular
decoherence rate $\gamma=\gamma_d+\gamma_\phi$. We note that while
molecular decay and dephasing as included here models exciton-phonon
interactions, it cannot represent exciton self-trapping, which can be
approximated by a nonlinear term in the energy functional
\cite{Rashba1994,Agranovich1999}. Self-trapping can lead to a further
reduction of the exciton propagation length in the weakly coupled
limit, but we have checked that it does not significantly affect the
strongly coupled limit. We thus neglect it in the following.

As mentioned above, we only include one cavity mode and describe the
molecules by a linear 1D chain along the longitudinal cavity direction
($x$ axis), with positions $\vec r_i = x_i \hat x$, such that the
cavity electric field is identical for all molecules. It is polarized
along the (out-of-plane) $z$ axis, leading to $\vec E_c(\vec
r_i)=E_c\hat z$. The total coupling between molecules and the cavity
mode can be characterized by the collective Rabi frequency $\Omega_R =
2\sqrt{\sum_i g_i^2}$. For zero detuning $\omega_m=\omega_c$, strong
coupling is entered for $\Omega_R>|\gamma-\kappa|/2$ and leads to
the formation of upper and lower polaritons at energies
$\omega_m\pm\frac12\sqrt{\Omega_R^2-|\gamma-\kappa|^2/4}$. The Rabi
splitting (energy difference between upper and lower polariton) can
approach $1~$eV in experiments \cite{Schwartz2011,Kena-Cohen2013}, and
can be tuned by changing either the molecule density or the mode
electric field strength.

In the following, we consider two types of molecular configurations: A
perfectly ordered distribution, with molecule positions on a regular
grid and dipole moments perfectly aligned to the electric field (i.e.,
along the $z$ axis), and a random distribution, where Gaussian noise
is added to the regular positions and the dipole moments are oriented
randomly. Note that for the present case of a 1D linear chain,
randomness is expected to suppress conductance much more efficiently
than in higher dimensions.

We next introduce a prescription for calculating an exciton
conductance $\sigma_e$, a steady-state quantity to characterize the
exciton transport efficiency similar to the electrical conductance for
charge transport. We assume that excitation is continuously pumped
into the system on the left side and measure the energy leaving the
system through the right side (cf.~\autoref{fig:system}). The pumping
is represented by an additional \emph{incoherent} driving term $\LL_p
= \gamma_p L_{\sigma_1^+}$.

The energy current is obtained from the rate of change of energy
\cite{Manzano2012}
\begin{equation}
\dot E=\frac{d}{dt}\langle H\rangle=\Tr(H\dot\rho) = 
\sum_\alpha \Tr(H\LL_\alpha[\rho]) \,.
\end{equation}
This is evaluated in the steady state, $\rho=\rho_{ss}$, for which the
total rate of change in energy is zero ($\dot\rho_{ss}=0 \to \dot
E=0$).  However, each of the Lindblad superoperators $\LL_\alpha$ can
be associated with a specific physical process. We thus identify the
energy current between the two reservoirs with the loss of energy from
the last molecule:
\begin{equation}\label{eq:current}
J = \gamma_d \Tr(H L_{\sigma_N^-}[\rho_{ss}]).
\end{equation}
We then define the exciton conductance as the current per driving
power, i.e., $\sigma_e = J/\gamma_p$ (which has units of energy).
Note that contrary to \cite{Manzano2012}, the energy entering the
system through pumping does not necessarily leave through the sink at
the end. It can also be lost through the radiative and nonradiative
decay of the molecules, as well as decay of the cavity mode.

We numerically obtain the steady state of the system using the
open-source QuTiP package \cite{Johansson2013}. To do so, we restrict
the total superoperator in \autoref{eq:master} to the zero- and
single-excitation subspaces. This truncation of the Hilbert space is
an excellent approximation in the presently relevant linear response
regime of weak pumping, i.e., for $\gamma_p$ smaller than the system
decay rates.

We choose the quantum emitter parameters to approximately correspond
to TDBC \emph{J} aggregates at room temperature
\cite{Moll1995,Valleau2012,Schwartz2013}: $\omega_m = 2.11~$eV,
$\gamma_r^{-1} = 500~$ps, $\gamma_{nr}^{-1} = 600~$fs, and
$\gamma_\phi^{-1} = 25~$fs. The cavity lifetime $\kappa^{-1} = 50~$fs
is typical for experiments using cavities made of thin metal mirrors
\cite{Orgiu2014}. The molecule parameters also determine the dipole
moment through $\gamma_r = \omega_m^3 d^2 / (3\pi \epsilon_0 \hbar
c^3)$, giving $d\approx 36~$D. The average intermolecular spacing
is taken as $\delta x=3~$nm. Note that while the coupling to the
cavity mode is taken into account explicitly, the radiative decay rate
into all other electromagnetic modes can also be modified by the
presence of a cavity. Since $\gamma_r$ is much smaller than the other
rates, this modification can safely be neglected here. While we only
show results for the parameters given above, we checked that the main
conclusions drawn in the following apply for a wide range of
parameters and do not depend on the specific values chosen here.

\begin{figure}[tbp]
  \includegraphics[width=\linewidth]{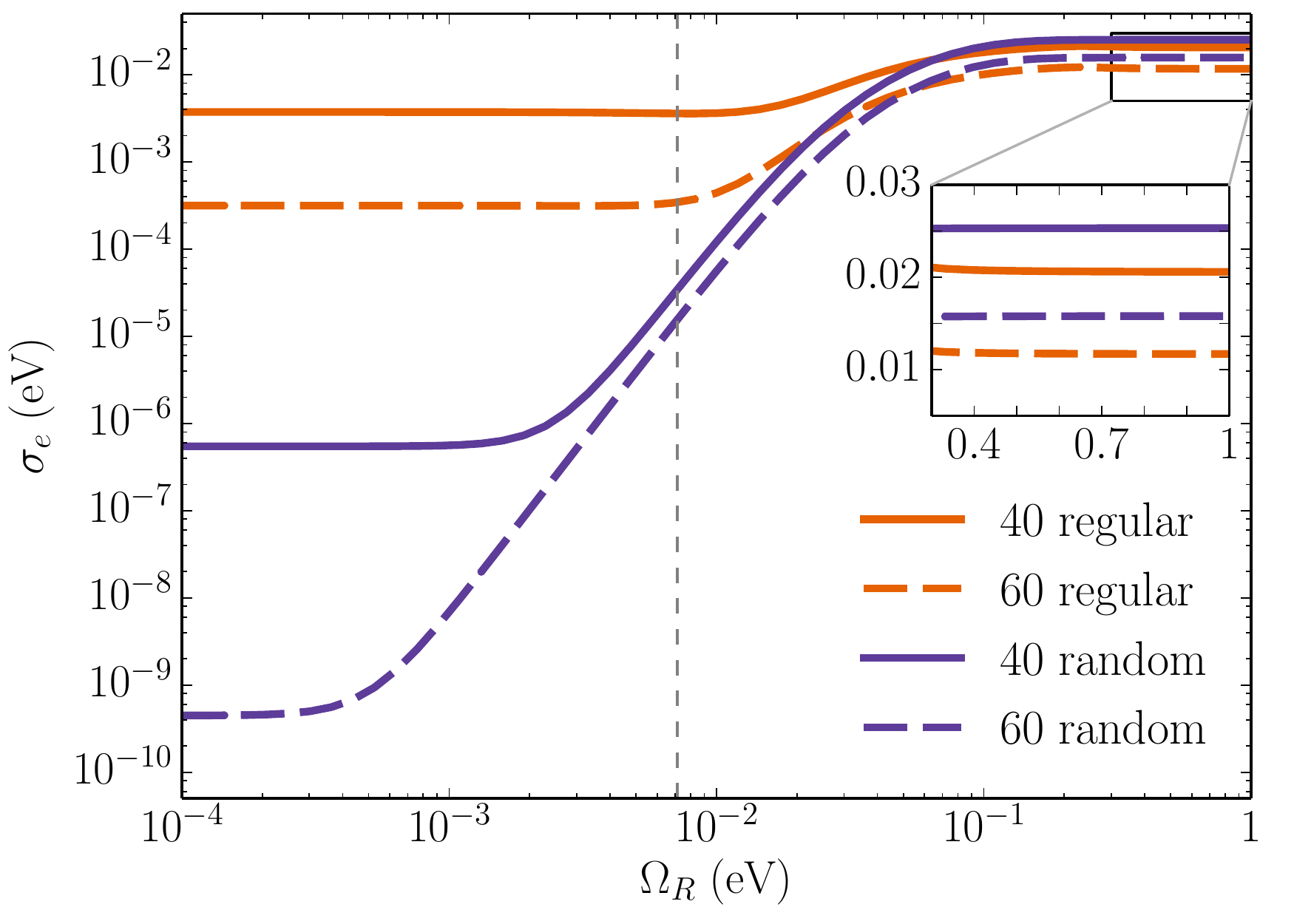}
  \caption{Exciton conductance at zero detuning as a function of the
    Rabi frequency for four different molecular configurations. The
    thin gray dashed line indicates the onset of strong coupling at
    $\Omega_R=|\gamma-\kappa|/2$. The number in the label indicates
    the number of molecules. The inset shows a zoom of the
    strong-coupling region.}
\label{fig:2}
\end{figure}

\autoref{fig:2} shows the exciton conductance $\sigma_e$ at zero
cavity-molecule detuning $\omega_c=\omega_m$, as a function of the
collective Rabi frequency $\Omega_R$. Here, we keep the number of
molecules fixed and change the electric field strength, going through
the transition from weak to strong coupling. This can be achieved in
an experiment by, e.g., putting the molecules at different positions
inside the cavity~\cite{Wang2014a}. We compare regular and random
molecule arrangements for chains of 40 and 60 molecules. For the
regular distribution, the strong dipole-dipole interaction leads to an
additional small energy shift $\Delta$ of the molecular bright state
coupling to the cavity; zero detuning thus corresponds to $\omega_c =
\omega_m + \Delta$.

For all of the cases shown in \autoref{fig:2}, the conductance is
approximately constant in the weak coupling limit $\Omega_R \ll
\gamma,\kappa$, where the cavity mode does not play a role.
Unsurprisingly, the conductance in this limit strongly depends on the
molecular configuration---it is almost completely suppressed for the
random case, for which 1D systems always show Anderson
localization. The conductance in the random case is calculated as the
logarithmic mean of 100 random configurations as appropriate
for localized systems, i.e., $\sigma_e =
\exp\langle\log\sigma_e^i\rangle$ \cite{Anderson1980}.
Note that even in the regular case, transport is quite inefficient due
to the relatively large decay and dephasing rates of the molecular
excitons, leading to diffusive transport \cite{Akselrod2014,Moix2013}.
Strikingly, when the coupling to the cavity mode is increased, an
extraordinary increase of the conductance is observed in all
cases. Once strong coupling is reached ($\Omega_R\gg \gamma,\kappa$),
the conductance again becomes almost independent of $\Omega_R$,
indicating that the fully formed polariton channel dominates exciton
conductance.  In this limit, the conductance also becomes almost
independent of the configuration and only depends on the number of
molecules, i.e., length of the 1D chain. While randomness can suppress
conduction almost completely in the weak-coupling limit, the polariton
modes are barely affected by it. As a consequence of their delocalized
nature induced by the collective exciton-cavity coupling, the
excitation can efficiently bypass the disordered chain of emitters.

This also provides a possible indication for the mechanism behind the
enhanced \emph{electrical} conduction observed under strong coupling
in the experiments by Orgiu \emph{et al.}~\cite{Orgiu2014}. However, the
connection between exciton transport through polaritons and electrical
conduction is currently unclear, as polaritons are, in principle,
neutral quasiparticles.

We also note that while we focus on incoherent driving for simplicity
in this work, we have found that under coherent driving or incoupling at
frequency $\omega$, the same general behavior is observed. The main
difference is an additional resonant enhancement when $\omega$
coincides with the eigenfrequencies of the system (see details in the
Supplemental Material \cite{supplemental}). An interesting aspect is
that even when hopping is completely suppressed, resonant transport
occurs not only when driving at the polariton eigenfrequencies, but
also at the unmodified molecule frequency---a clear signature that the
dark states (which are not coupled to the cavity EM field) are still
affected by the existence of strong coupling.

\begin{figure}[tbp]
\includegraphics[width=\linewidth]{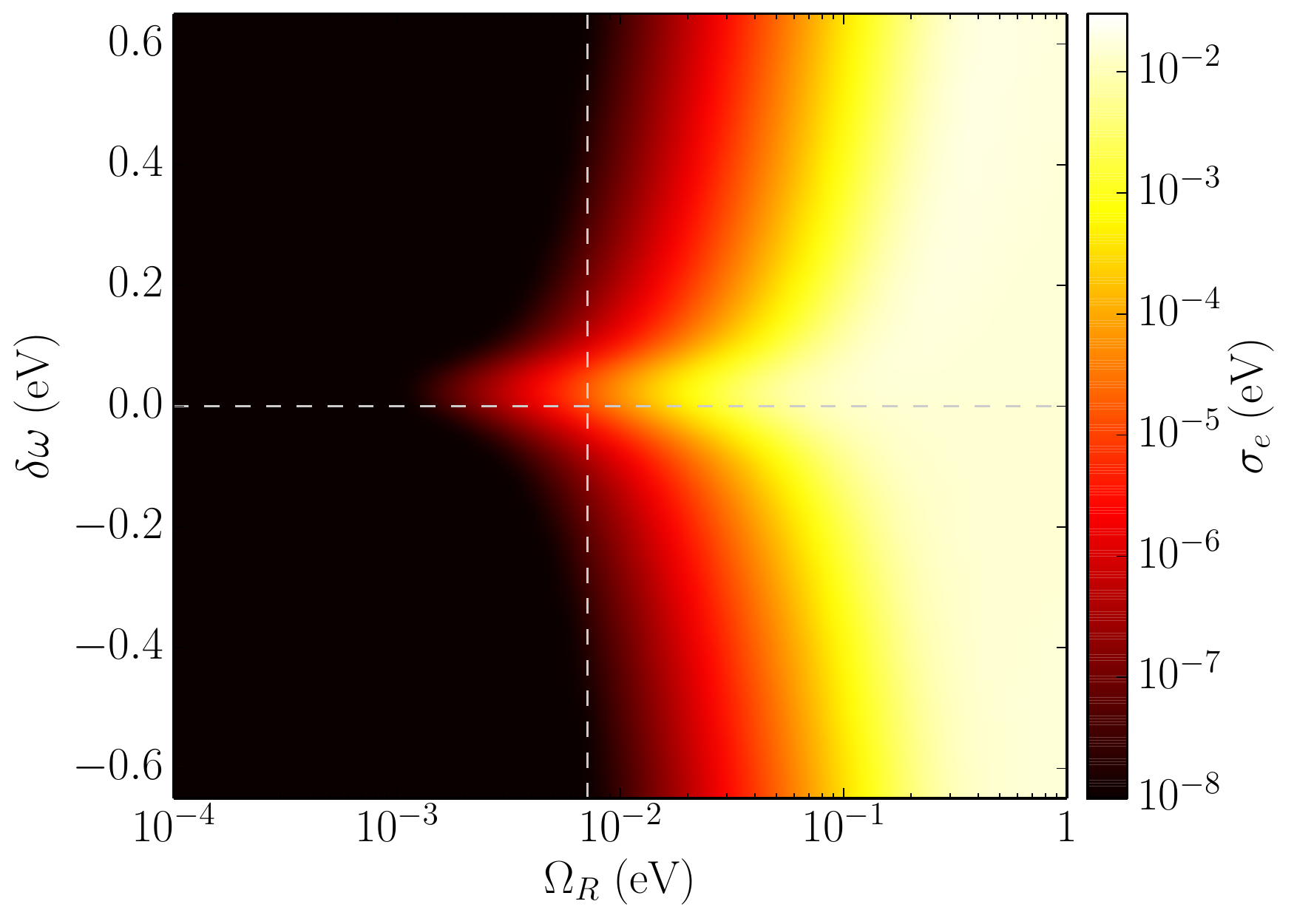}
\caption{Conductance as a function of Rabi splitting and detuning for
  random configurations of 60 emitters (averaged over 100 random
  realizations). The thin gray dashed lines indicate $\delta\omega=0$
  and the onset of strong coupling at $\Omega_R=|\gamma-\kappa|/2$.}
\label{fig:gscan_detuning}
\end{figure}

We next focus on the case where the cavity mode is detuned from the
molecular excitations by an energy $\delta\omega = \omega_m -
\omega_c$. As shown in \autoref{fig:gscan_detuning}, the onset of the
extraordinary conductance is then shifted to larger coupling strengths
for increasing detuning $|\delta\omega|$. However, the final
conductance in the strong-coupling limit is independent of the
detuning. This again indicates that the conduction proceeds through
the polariton modes, which are only fully formed when the Rabi
frequency $\Omega_R$ becomes large enough to not only overcome
decoherence processes, but also the detuning. Once this is fulfilled,
their character does not strongly depend on the detuning. However,
even for relatively large and experimentally relevant Rabi
frequencies, e.g., $\Omega_R=100~$meV, a small change of the detuning
can strongly suppress or enhance conductance. If the detuning could be
modified dynamically in an experiment (e.g., by displacing a cavity
mirror with a piezo), this would enable novel applications based on
switching of the exciton conductance.

Next, we develop a simplified model to understand the extraordinary
increase in exciton conductance under strong coupling. The main idea
behind it is that there are two almost independent transport channels:
(i) Direct excitonic transport through hopping between the molecules,
which dominates in the weak-coupling limit, and (ii) polaritonic
transport through the collective modes created by strong coupling to
the cavity, which increases rapidly (polynomially) as the coupling is
increased and saturates in the strong-coupling limit. To expose the
contribution of the second channel unambiguously, we remove the
dipole-dipole interaction responsible for hopping from the Hamiltonian
\autoref{eq:hamiltonian}. We furthermore assume that the molecules are
all aligned along the cavity field polarization axis $z$. The
interaction with the cavity mode is then identical for all molecules.

\begin{figure}[tbp]
\includegraphics[width=\linewidth]{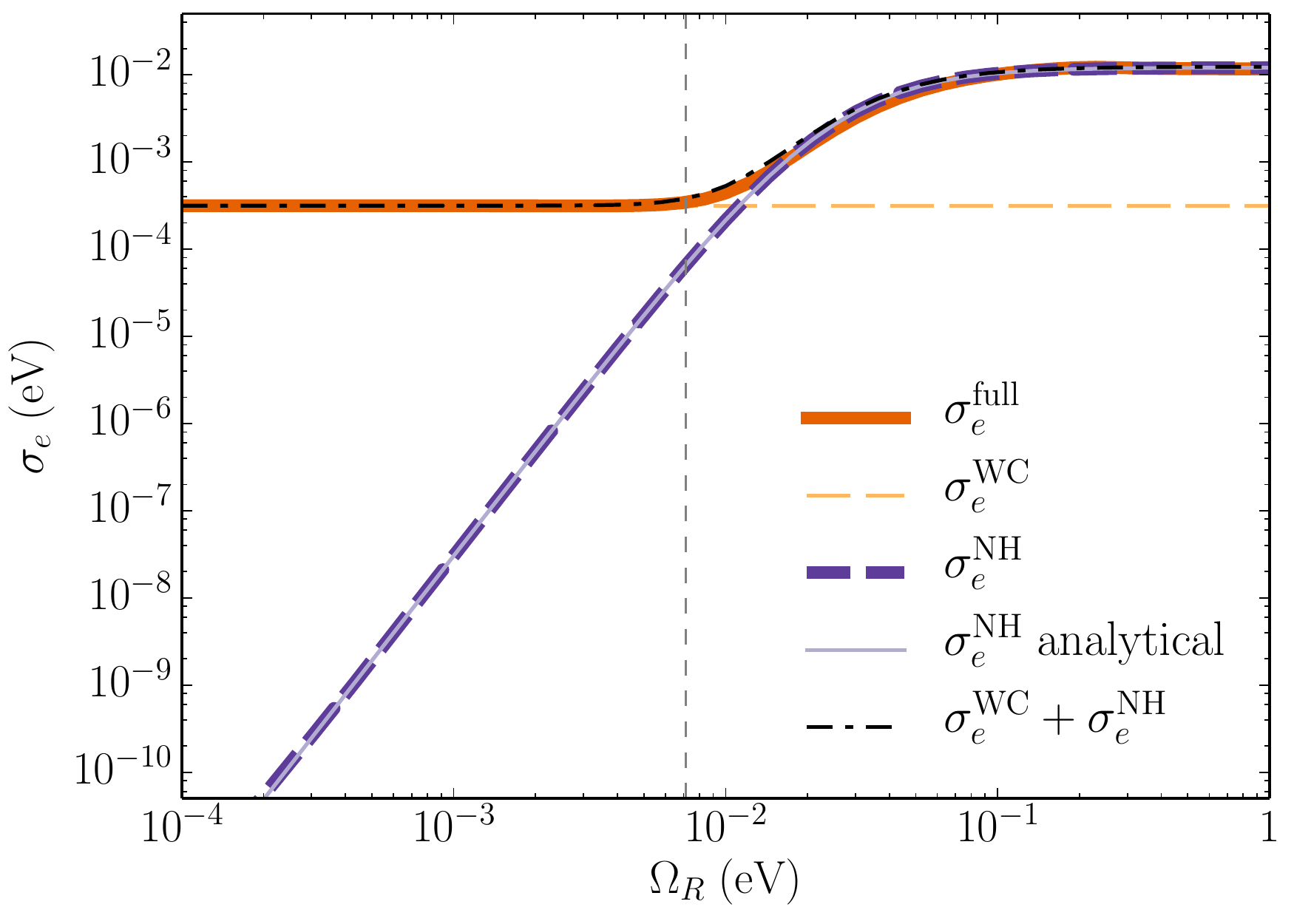}
\caption{Transmission for a regular chain of 60 emitters. The solid
  orange line gives the results for the full model, with its
  weak-coupling limit $\sigma_e^{\mathrm{WC}}$ indicated by the thin
  dashed orange line. The result without hopping (dashed purple line),
  i.e., with dipole-dipole interactions turned off, is perfectly
  reproduced by the analytical result \autoref{eq:condmod_nodw} (thin
  light purple line). The dash-dotted black line is the sum of the two
  independent transport channels of direct hopping in the
  weak-coupling limit and the cavity-mediated contribution without
  hopping (see text). The vertical gray dashed line again indicates
  the onset of strong coupling, $\Omega_R=|\gamma-\kappa|/2$.}
\label{fig:model}
\end{figure}

\autoref{fig:model} shows that the picture of independent channels is
indeed valid: For the conductance $\sigma_e^{\mathrm{NH}}$ without
hopping, the transmission plateau for weak coupling disappears, while
the transmission in strong coupling is essentially unchanged. The
polariton contribution without hopping decreases rapidly with
decreasing Rabi frequency. The independence of the two transport
channels is further verified by plotting the sum of the exciton
conductance in the weak-coupling limit,
$\sigma_e^{\mathrm{WC}}\!=\!\sigma_e^{\mathrm{full}}(\Omega_R\!=\!0)$,
and the cavity-mediated contribution
$\sigma_e^{\mathrm{NH}}(\Omega_R)$ without hopping. This sum, given by
the dash-dotted black line in \autoref{fig:model}, agrees excellently
with the full result, indicating that the two transport channels are
indeed effectively independent.

Since in this model, all molecules but the first (due to the pumping)
are indistinguishable, they behave identically. Independent of the
number of molecules $N$, there are thus just 12 independent components
of the steady state density matrix within the zero- and
single-excitation subspace. This density matrix can then be obtained
analytically, and inserting the solution into \autoref{eq:current}
gives an explicit formula for the conductance. In the linear response
limit of weak driving and for zero detuning $\omega_c=\omega_m$, it is
given by
\begin{widetext}
\begin{equation}\label{eq:condmod_nodw}
\sigma_e = \frac{\gamma \gamma_d (\gamma+\Gamma) \omega_m \Omega_R^4}
{\left(     2\gamma\gamma_d\Gamma N + ( 2    \gamma_\phi + \gamma_d       N) \Omega_R^2 \right)
 \left(\kappa\gamma\gamma_d\Gamma N + (\kappa\gamma_\phi + \gamma_d\Gamma N) \Omega_R^2 \right)}
\end{equation}
\end{widetext}
where we have defined the total decoherence rate $\Gamma = \gamma +
\kappa$. More details and the full expression for arbitrary detuning
are given in the Supplemental Material \cite{supplemental}.

As expected, the analytical solution (\autoref{eq:condmod_nodw})
perfectly matches the numerics (cf.~\autoref{fig:model}). For small
Rabi frequency, the conductance through the polariton modes grows with
$\Omega_R^4$, while it saturates to a constant value for
$\Omega_R\to\infty$. For large $N$, this constant value is given by
$\gamma\omega_m/(\gamma_d N^2)$ if $\kappa\gg\gamma$ and by
$2\gamma_\phi\omega_m/(\gamma_d N^2)$ if
$\gamma_\phi\gg\kappa,\gamma_d$. Importantly, the decay with system
size is algebraic $(N^{-2})$, as opposed to the localized $e^{-N}$
behavior expected in the absence of strong coupling. Note that $N$
occurs here because only a single molecule is connected to each of the
baths; for a quasi-1D wire with a transverse extension, the length of
the system would be the relevant variable. The dependence on the
fourth power of $\Omega_R$ in the weak-coupling limit is explained by
the rate obtained from two quantum jumps, to the cavity mode and back,
with coupling $\propto\Omega_R$. Interestingly,
\autoref{eq:condmod_nodw} shows that the conductance, at least in the
simplified model without disorder or direct hopping, is not directly
related to the conventional criterion for the onset of strong coupling
where the vacuum Rabi splitting becomes real ($\Omega_R >
|\gamma-\kappa|/2$). Indeed, the difference $\gamma-\kappa$ does
not occur in \autoref{eq:condmod_nodw}. Instead, the exciton
conductance becomes constant when $\Omega_R^2\gg2\gamma\Gamma$ (for
large $N$). This is related to large values of the
\emph{cooperativity} $C=\Omega_R^2/\gamma\kappa$, which can occur even
if strong coupling is not fully reached.

To conclude, we have demonstrated that the formation of polariton
modes, i.e., strong coupling, can dramatically enhance exciton
transport. When the coupling is strong enough and the polaritons are
fully formed, the excitons can almost completely bypass the chain of
quantum emitters and ``jump'' directly from one end to the other,
leading to large exciton conductance. This robust effect persists
almost independently of the exact parameters of the system, and most
notably occurs efficiently even when the underlying excitonic system
is strongly disordered and its transport is completely suppressed due
to localization. Through a simple model, we have furthermore shown
that transport through direct hopping and through the polariton modes
constitute two effectively independent channels, which helps to
explain why the polariton conductance is almost independent of the
disorder in the system.  These results demonstrate a possible pathway
for improving the efficiency of excitonic devices, where the EM mode
could be provided by plasmonic structures to enable fully integrated
nanometer-scale devices. We note that related results have
simultaneously been obtained by Schachenmayer \emph{et al.}~\cite{Schachenmayer2014}.

\begin{acknowledgments}
  We thank T.~Ebbesen, G.~Pupillo, C.~Genes, and L.~Mart\'in Moreno for
  fruitful discussions, and E.~Moreno and A.~Delga for a careful
  reading of the manuscript. This work has been funded by the European
  Research Council (ERC-2011-AdG Proposal No. 290981).
\end{acknowledgments}

\bibliography{extranotes,mendeley}

\newpage
\begin{widetext}

\begin{center}
\textbf{\textsc{\Large Supplemental Material}}
\end{center}

\section{Coherent driving}
In this section, we show some results for coherent driving of the
system (as opposed to the incoherent driving used in the main
text). As mentioned in the main text and shown below, the main
conclusions are similar to those obtained under incoherent
driving. The main difference is that under coherent driving at
frequency $\omega$, a resonant enhancement of the conductance is
observed when $\omega$ corresponds to one of the eigenmodes of the
system. We use the same model as in the main text, but instead of an
incoherent pumping term described by $\LL_p$, we add a coherent
driving term $H_p(t) = \Omega_p \sigma_1^- e^{i\omega t} + c.c.$
(within the rotating wave approximation) to the Hamiltonian. Here,
$\Omega_p$ and $\omega$ are the driving amplitude and frequency,
respectively. The steady state can then be calculated by going to the
rotating frame, in which $H_p$ is time-independent, but the system
frequencies are shifted by $\omega$. The exciton conductance is
calculated analogously to the incoherent case, using
$\sigma_e^{\mathrm{coh}} = \hbar J/\Omega_p^2$, where the factor
$\hbar$ is added to make $\sigma_e^{\mathrm{coh}}$ unitless.

\begin{figure}[bp]
  \includegraphics[width=\linewidth]{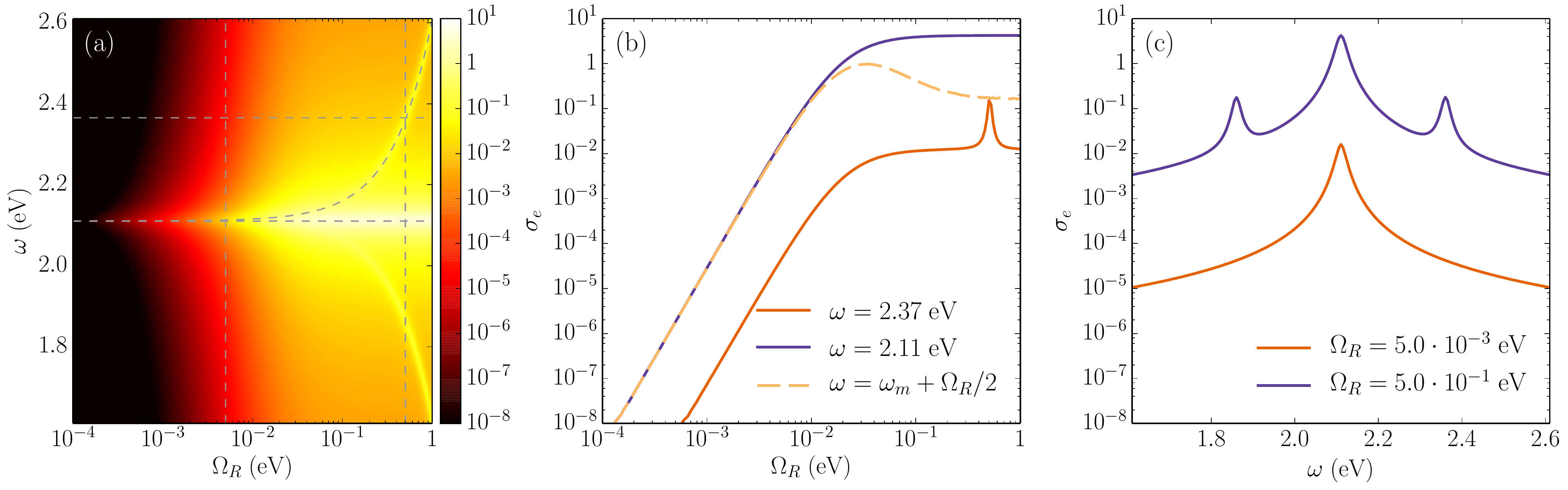}
  \caption{Exciton conductance as a function of Rabi splitting and
    driving frequency for a chain of $20$ molecules, where the
    dipole-dipole interaction has been suppressed to only show the
    polariton contribution to conductance. Panel (a) shows a map of
    the conductance as a function of total Rabi frequency $\Omega_R$
    and driving frequency $\omega$, while the other two panels show
    cuts of the data at different driving frequencies (b) and at two
    Rabi frequencies (c). The positions of the cuts are indicated by
    the dashed grey lines in (a).}
\label{fig:coh_driving}
\end{figure}

We focus again on the case without dipole-dipole interaction, which
allows to calculate only the polariton contribution to exciton
conductivity even when relatively short chains are used ($N=20$ in the
following). For simplicity, we also choose zero cavity-molecule
detuning, $\omega_c=\omega_m$. \autoref{fig:coh_driving} demonstrates
that the polariton-mediated exciton conductance under coherent driving
displays a polynomial increase with $\Omega_R$, which saturates to a
constant value once strong coupling is fully entered. This is
identical to the behavior under incoherent pumping. In addition, the
conductance is resonantly enhanced when the driving frequency
coincides with an eigenfrequency of the system. The system
eigenfrequencies correspond to the upper and lower polariton states
($\omega_m\pm\Omega_R/2$), as well as the dark states at $\omega_m$,
which are not directly coupled to the cavity. Interestingly, and
somewhat surprisingly, the largest conductance is found when driving
the dark modes ($\omega=\omega_m$), \emph{not} the polariton modes
(cmp.\ the violet and dashed orange lines in
\autoref{fig:coh_driving}b). This effect persists even when dephasing
(which couples between dark and bright states) is turned off, and
demonstrates that even the eigenstates that are not themselves
strongly coupled to the cavity mode become delocalized when the system
enters strong coupling.

\section{Simplified model}
In this section, we provide more details on the simplified model and
its solution described in the text. The starting point is the
Hamiltonian given in Eq.~1 in the main text:
\begin{equation}\label{eq:hamiltonian}
H = \omega_c a^\dag a + \sum_i \omega_m \sigma_i^+ \sigma_i^- 
+ \sum_i g_i (a^\dag\sigma_i^- + a\sigma_i^+)
+ \sum_{i,j} V^{dd}_{ij} (\sigma_i^+ \sigma_j^- + \sigma_j^+\sigma_i^-) \,,
\end{equation}
from which we remove the direct dipole-dipole interaction responsible
for hopping by setting $V^{dd}_{ij}=0$. We next assume that all
molecules are oriented identically, such that the $g_i$ are constant
and given by $g_i=\Omega_R/\sqrt{4N}$. As explained in the main text,
all molecules apart from the first one, which is pumped, are now
indistinguishable. The time evolution of the system density matrix is
given by Eq.~2 in the main text:
\begin{equation}\label{eq:master}
\dot\rho = -\frac{i}{\hbar} [H,\rho] + \sum_{\alpha=i,c} \LL_\alpha[\rho] \,.
\end{equation}
Due to the structure and symmetries of the system, we can write the
steady-state density matrix within the zero- and single-excitation
subspace as
\begin{equation}
\rho_{ss} = \left(\begin{array}{ccccccc}
 \rho_{00} & 0 & 0 & 0 & 0 & 0 & 0 \\
 0 & \rho_{cc}   & \rho_{c1}   & \rho_{c2}   & \hdotsfor{2}           & \rho_{c2} \\
 0 & \rho_{c1}^* & \rho_{11}   & \rho_{12}   & \hdotsfor{2}           & \rho_{12} \\
 0 & \rho_{c2}^* & \rho_{12}^* & \rho_{22}   & \rho_{23}  & \hdots    & \rho_{23} \\
 0 & \trvdots   &  \trvdots   & \rho_{23}^* & \sddots    & \sddots    & \svdots   \\
 0 &            &             & \svdots    & \sddots    & \sddots    & \rho_{23}  \\
 0 & \rho_{c2}^* & \rho_{12}^* & \rho_{23}^* & \hdots     & \rho_{23}^* & \rho_{22} \\
\end{array}\right) \,,
\end{equation}
where $\rho_{00}$ is the probability to be in the ground state (no
photons or excitons), $\rho_{cc}$ is the cavity excitation
probability, $\rho_{ii}$ is the excitation probability of molecule
$i$, and $\rho_{ab}$ is the coherence matrix element between
constituent $a$ and $b$. Since all but the first molecule behave
identically, the density matrix elements involving molecule $i$ with
$i>1$ are all identical and given with $i=2$. Additionally,
$\rho_{23}$ denotes the coherence element between \emph{any} pair of
molecules not involving the first.

The steady state density matrix can then be obtained from
$\dot\rho_{ss}=0$, which for the specific form of the density matrix
corresponds to a set of linear equations for the 4 real probabilities
$\{\rho_{00}, \rho_{cc}, \rho_{11}, \rho_{22}\}$ and the 4 complex
coherence elements $\{\rho_{c1}, \rho_{c2}, \rho_{12}, \rho_{23}\}$,
giving 12 independent real variables. Note that there are actually
$(N+1)^2+1$ linear equations for $N$ molecules. However, due to the
symmetry, only 11 of them are linearly independent. The steady-state
density matrix is then obtained by solving for the null space of the
coefficient matrix of the linear equations, and subsequently
normalized to $\Tr\rho_{ss}=1$.

Having obtained the steady state density matrix, the exciton
conductivity is calculated using Eq.~7 in the main text, which
explicitly gives
\begin{equation}
\sigma_e = \frac{\gamma_d}{\gamma_p} 
\left(\frac{\Omega_R}{2\sqrt{N}} \Re(\rho_{c2}) + \omega_m \rho_{22}\right) \,.
\end{equation}
Inserting the solution $\rho_{ss}$ into this expression and taking the
linear-response limit $\gamma_p\to0$ gives the general solution for
$\sigma_e$, which we give here for completeness:
\newcommand{\dw}{\delta\omega}
\newcommand{\gf}{\gamma_\phi}
\newcommand{\gd}{\gamma_d}
\newcommand{\ka}{\kappa}
\newcommand{\ga}{\gamma}
\newcommand{\Ga}{\Gamma}
\begin{equation}
  \sigma_e = \frac{\ga\gd\Omega_R^4 (4N\ga ((\ka^2\ga + 2\ka(\ga^2+\Delta^2) + \ga^3)\omega_m - \ga\gd\Ga\dw) + ((2\gf+N\gd) \ka\dw + N(\ga+\Ga) \Ga \omega_m) \Omega_R^2)}
  {N (16N\ga^2\gd\Delta^2 + 4(\gf+N\gd)\ga\Ga\Omega_R^2 + (2\gf+N\gd) \Omega_R^4) (4N\ka\ga\gd\Delta^2 + (\ka\gf+N\Ga\gd) \Ga \Omega_R^2)} \,,
\end{equation}
where in addition to the definitions in the main text, we use
$\Delta^2 = \dw^2 + \Ga^2/4$. Finally, we note that the same approach
works equally well under coherent driving (not shown here).
\end{widetext}

\end{document}